\documentclass[journal=jacsat,manuscript=article,layout=twocolumn]{achemso}

\usepackage{mathrsfs}
\usepackage[scr=rsfso]{mathalpha}
\usepackage{fontawesome5}
\usepackage{pifont}
\usepackage{multirow}%
\usepackage{amsmath,amssymb,amsfonts}%
\usepackage{amsthm}%
\usepackage[title]{appendix}%
\usepackage{manyfoot}%
\usepackage{booktabs}%
\usepackage{algorithm}%
\usepackage{algorithmicx}%
\usepackage{algpseudocode}%
\usepackage{listings}%
\usepackage{pdfpages}%

\usepackage[version=3]{mhchem} 
\usepackage[utf8]{inputenc}
\usepackage[english]{babel}
\usepackage{lmodern}
\usepackage{hyperref}
\usepackage{enumitem}
\usepackage{chemfig}
\usepackage{pifont}
\usepackage{verbatim}
\usepackage{adjustbox}
\usepackage{tikz}
\usepackage{chemgreek}
\usepackage{upgreek}
\usepackage{csquotes}
\usepackage{bm}

\DeclareUnicodeCharacter{2212}{\ensuremath{-}}

\usetikzlibrary{calc}
\usetikzlibrary{arrows.meta}
\pgfmathsetseed{1034}

\tikzset{
  pics/water/.style args={#1,#2}{code={
    \pgfmathsetmacro{\rotC}{random()*360}
    \def\rot{\rotC+230}
    \def\sc{#2*0.9}

    \begin{scope}[rotate=\rot+89,scale=\sc]
    \node[rotate=\rot, scale=\sc, inner sep=0
    ] at (0,0.01)
        {\includegraphics[width=1.5cm]{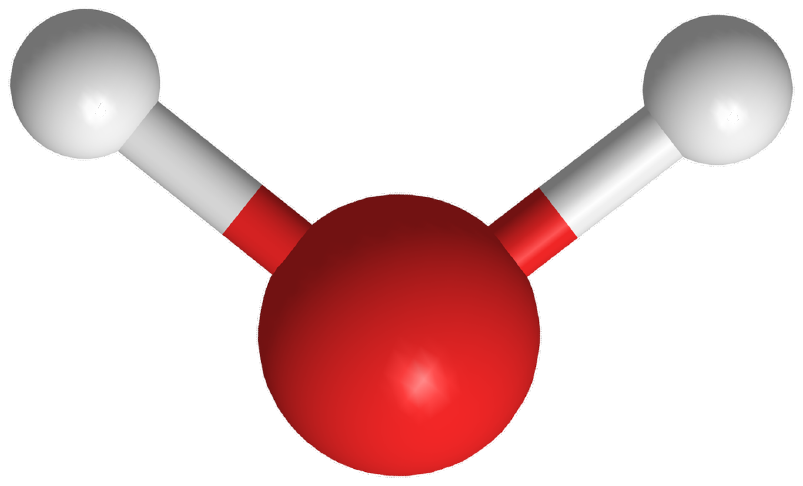}};
        \draw[->,thick,color=ForestGreen!70!black] (0.4*\sc,0) -- (-0.25*\sc,0);
    \end{scope}
  }}
}

\newcommand*{\mrmu}{\bm{\mathrm\mu}}
\newcommand*{\tdd}{\mathrm{d^*}}
\newcommand*{\metricd}{\mathcal{D}}

\newcommand*{\correc}[1]{#1}

\definecolor{beige}{HTML}{fee8c8}
\definecolor{ocre}{HTML}{fdbb84}
\definecolor{braise}{HTML}{e34a33}

\definecolor{arrowred}{HTML}{ff3e3e}
\definecolor{arroworange}{HTML}{ff964f}
\definecolor{arrowblue}{HTML}{1982c4}

\author{Leon Huet}
\affiliation[CMSP, ICTP]{Condensed Matter and Statistical Physics, The Abdus Salam International Centre for Theoretical Physics, Strada Costiera 11, Trieste 34151, Italy}
\email{lhuet@ictp.it}
\author{Vittorio Del Tatto}
\affiliation[SISSA]{Scuola Internazionale Superiore di Studi Avanzati, Trieste 34136, Italy}
\author{Debarshi Banerjee}
\affiliation[CMSP, ICTP]{Condensed Matter and Statistical Physics, The Abdus Salam International Centre for Theoretical Physics, Strada Costiera 11, Trieste 34151, Italy}
\alsoaffiliation[SISSA]{Scuola Internazionale Superiore di Studi Avanzati, Trieste 34136, Italy}
\author{Alessandro Laio}
\affiliation[CMSP, ICTP]{Condensed Matter and Statistical Physics, The Abdus Salam International Centre for Theoretical Physics, Strada Costiera 11, Trieste 34151, Italy}
\alsoaffiliation[SISSA]{Scuola Internazionale Superiore di Studi Avanzati, Trieste 34136, Italy}
\author{Ali A. Hassanali}
\affiliation[CMSP, ICTP]{Condensed Matter and Statistical Physics, The Abdus Salam International Centre for Theoretical Physics, Strada Costiera 11, Trieste 34151, Italy}
\email{ahassana@ictp.it}

\title[Water IB]{Causality in Liquid Water as a Hallmark of Emergent Glassy Dynamics}

\abbreviations{IR,NMR,UV}

\begin{document}

\begin{abstract}

In molecular liquids such as water, time-delayed influences between microscopic or mesoscopic variables
are typically probed using time-correlation functions, which are symmetric under detailed balance and therefore blind to dynamical asymmetries. Here, we characterize water’s dynamics using a causal inference metric that captures asymmetric couplings between collective variables. Analyzing equilibrium molecular dynamics simulations at ambient conditions and in the high-density liquid (HDL) regime of supercooled water, we uncover pronounced asymmetries in the couplings  between orientational and translational degrees of freedom across multiple time and length scales. \correc{At room temperature, rotational modes exert a modest directional influence on translational dynamics} 
In contrast, in the supercooled HDL regime, translational motions emerge as the primary drivers of the dynamics, suggesting facilitation-like relaxation mechanisms characteristic of glassy systems. These results reveal a qualitative reorganization of dynamical couplings across thermodynamic conditions, implying that  molecular liquids at thermal equilibrium can exhibit an emergent directionality in their fluctuation couplings. As a consequence, our analysis reveals that external perturbations acting on specific degrees of freedom can induce a stronger arrow of time in the causal relations between translational and orientational modes.

\end{abstract}

\section{Significance Statement}

Understanding how molecules collectively move in liquids is essential for explaining phenomena ranging from solvation to glass formation. Conventional analyses rely on correlation functions, which cannot distinguish directional influences between molecular motions. Here, we apply a causal inference framework to liquid water and uncover hidden asymmetries between locally resolved translational and orientational dynamics within water solvation shells. Under ambient conditions, these motions 
\correc{display a weak coupling}, whereas supercooling induces a qualitative shift in which translational rearrangements drive molecular reorientation, consistent with facilitation mechanisms in glassy materials. More broadly, our approach suggests experimentally testable routes to probe directional couplings in liquids.

\section{Introduction}

Liquid water is a ubiquitous solvent, covering ~70\% of Earth’s surface, driving weather cycles, and serving as the primary medium of life~\cite{ball_water_2008,bellissent-funel_water_2016}. It lies at the center of many physical and chemical processes, from ion transport and chemical activation to protein folding~\cite{ando_hf_1995,mallamace_possible_2011,liu_observation_2005,choudhury_dewetting_2007,marcus2006}. Despite its molecular simplicity, water exhibits an unusually rich phase diagram and a range of thermodynamic and dynamical anomalies~\cite{lang_anomalies_1982}, including a density maximum~\cite{kell_density_1975, baez_existence_1994} and a non-Arrhenius evolution of the dynamics~\cite{lang_anomalies_1982, stirnemann_communication_2012,Ito1999,kimmel2021}. These distinctive properties are widely attributed to the fluctuating structure and dynamics of water’s hydrogen-bond network (HBN)~\cite{kontogeorgis_water_2022,fecko_local_2005}.

It is well established that molecular motion in liquid water involves cooperative reorganization of the HBN, yet the spatial extent of these processes and their evolution
across the phase diagram remain open questions. Laage and Hynes demonstrated through molecular simulations that water’s rotational dynamics proceed via large angular jump motions rather than small diffusive steps~\cite{laage_molecular_2006,laage_water_2009,laage_water_2012}. Subsequent studies have shown that these reorientation events require collective rearrangements, which couple density fluctuations and hydrogen-bond coordination defects\cite{liu_correlation_2013,liu_interplay_2018,offei-danso_collective_2023}. A central challenge in rationalizing these mechanisms lies in the high dimensionality of the fluctuation space~\cite{ohmine_liquid_1995} and in identifying suitable statistical measures to probe the underlying dynamics.

In molecular systems, the coupling between two collective variables (CVs) $\mathbf{A}(t)$ and $\mathbf{B}(t)$,
is typically quantified through lagged cross-correlations $\mathbb{E}\left[\mathbf{A}(0)\cdot \mathbf{B}(\tau) \right]$.
For molecular liquids, cross-correlations between dipoles ($\bm\mu_i$ and $\bm\mu_j$) or polarizabilities ($\bm\alpha_i$ and $\bm\alpha_j$) form the basis of linear-response functions underlying spectroscopies such as infrared (IR)~\cite{guillot_molecular_1991,silvestrelli_ab_1997, scherlo_ir_2025}, Raman~\cite{auer_ir_2008, mcquarrie_statistical_1975,galli2013} and dielectric spectroscopy~\cite{mcquarrie_statistical_1975,arbe2016}. 

When the dynamics is time-reversible, these correlation functions are symmetric under the exchange of the two CVs~\cite{del_tatto_towards_2025}.
In these conditions, response functions written in terms of cross-correlations via fluctuation-dissipation relations cannot probe directional couplings, which underlie cause-effect relationships.

\correc{In contrast, many descriptions of aqueous molecular processes implicitly assume directionality between coupled degrees of freedom. Descriptions of immobility in liquids, commonly referred to as caging effects under supercooled conditions, invoke dynamic facilitation, whereby structural relaxation enables subsequent molecular motion \cite{garrahan_coarse-grained_2003,chandler_dynamics_2010,gokhale_growing_2014,biroli_perspective_2013,ortlieb_probing_2023,saito_unraveling_2024,donati_stringlike_1998,keys_excitations_2011}. Similarly, within Marcus theory of electron transfer, solvent reorganization is treated as the collective coordinate governing barrier crossing, with translational and rotational solvent rearrangements contributing to the solvent reorganization energy. More generally, electron and proton transfer reactions are frequently interpreted in terms of solvent motions or electric-field fluctuations modulating the reactive coordinate and facilitating charge transfer ~\cite{saura_electric_2022, hassanali_proton_2013, siwick_role_2007}. Solvent fluctuations have likewise been proposed to facilitate conformational transitions in proteins \cite{rhee_solvent_2008, frauenfelder_protein_2006, lucent_protein_2007} and to promote hydrophobic collapse during protein aggregation through dewetting transitions \cite{zhou_hydrophobic_2004,liu_observation_2005,athawale_effects_2007}.} 

\correc{On pure water dynamics specifically, previous studies have established that translational and rotational motions in supercooled water become increasingly coupled upon cooling. Within the framework of Mode-Coupling Theory (MCT)\cite{bengtzelius_dynamics_1984, reichman_mode-coupling_2005}, this coupling is understood as a consequence of transient caging: molecular rotations remain constrained by their local environment until structural relaxation enables both translational displacements and rotational reorientation. At lower temperatures, where activated hopping becomes increasingly important, translational cage escape  facilitates rotational relaxation. Correlation-based analyses have provided compelling evidence for this picture. For example, Chen et al.\cite{chen_molecular-dynamics_1997} demonstrated a growing coupling between translational and rotational dynamics upon supercooling through connected translational-rotational correlation functions, while Faraone et al.\cite{faraone_model_2003} characterized this coupling using joint translational-rotational probability distributions that are directly accessible in neutron scattering experiments. Together, these studies established the intimate relationship between translational and rotational relaxation in supercooled water.}

\correc{These examples illustrate a common paradigm in chemical physics: directional influence is frequently inferred from mechanistic models (for example Marcus theory) or the separation of characteristic timescales. Such paradigms have proven remarkably successful in predicting experimental observations, yet they are \emph{interpretations}, and not approaches which can be used to infer the  directionality of dynamical influence  from molecular trajectories. Traditional analyses based on time-correlation functions and relaxation times characterize the strength and timescales of coupled fluctuations, but because these observables are  symmetric under equilibrium dynamics satisfying detailed balance, they cannot determine which degree of freedom is dynamically leading and which is following. All these examples raise a fundamental question: is there a quantitative framework capable of rigorously identifying cause-and-effect relationships in aqueous molecular systems?}

\correc{To address this question, rather than quantifying the strength of translational-rotational coupling through statistical correlations, we employ a causal inference method to infer the directional influence between translational and rotational dynamics.}

Recently, we investigated causality in molecular systems undergoing equilibrium and time-reversible dynamics, using causal inference measures designed to detect asymmetric dynamical couplings~\cite{del_tatto_towards_2025}.
In the causal inference literature, a widely used framework to quantify directed dependencies between time-dependent variables $\mathbf{A}(t)$ and $\mathbf{B}(t)$ is Granger causality (GC)~\cite{granger_investigating_1969,barrett_multivariate_2010,blinowska_granger_2004}, which evaluates whether the prediction of the future state of $\mathbf{B}$, based on its own past, can be improved by including information from the past of $\mathbf{A}$. GC and related approaches~\cite{schreiber_measuring_2000,del_tatto_robust_2024} allow for identifying causal links by assuming~\emph{causal sufficiency}, namely that all common drivers of $\mathbf{A}$ and $\mathbf{B}$ are included in the analysis~\cite{runge_causal_2018}. Applied at the pairwise level, this implies that there is no common variable driving both $\mathbf{A}$ and $\mathbf{B}$. More generally~\cite{spirtes_causation_2001}, the absence of a causal link from $\mathbf{A}$ to $\mathbf{B}$ remains valid even when inferred in a pairwise estimation. 

The presence of causal directed links between different variables can be interpreted in  straightforward experimental terms: it predicts how manipulating a particular variable—even beyond small perturbations—will affect the rest of the system. A link from $\mathbf{A}$ to $\mathbf{B}$ means that manipulating $\mathbf{A}$ will alter the future evolution of $\mathbf{B}$, while the absence of a link means that intervening on $\mathbf{A}$ leaves $\mathbf{B}$’s future state unchanged. In molecular systems satisfying detailed balance, as discussed in Ref.~\cite{del_tatto_towards_2025}, the existence of a causal link between two CVs is inherently linked to the time scale at which such a causal effect manifests.

In this work, we take a step toward addressing causal questions for dynamical fluctuations in liquid water under different thermodynamic conditions. In particular, we investigate signatures of causal relationships
among orientational and translational degrees of freedom,
both under ambient and in the high-density liquid (HDL) phase of supercooled water.
Our analysis relies on a statistical measure called Imbalance Gain (IG)~\cite{del_tatto_robust_2024} (reviewed in \hyperref[section:Method]{Methods}), which extends the predictive criterion of GC to high-dimensional variables. 
\correc{This approach allows us to identify not only the emergence and strengthening of translational-rotational coupling upon supercooling, consistent with previous studies, but also its directional asymmetry.}

\section{Results}

\begin{figure}[t]
    \centering
    \includegraphics[width=\linewidth]{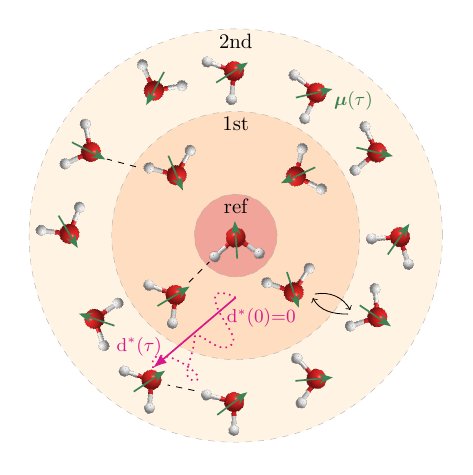}
    \vspace{-3em}
    \caption{Schematic depiction of the different variables in our local water environment.
    The shells' boundaries are fixed at 3.3~\AA~and 5.7~\AA. $\mrmu_{\mathrm{ref}}$ is the central dipole moment, $\mrmu_{\mathrm{1st}}$ and $\mrmu_{\mathrm{2nd}}$ are the sums of the dipoles in the first and second shells, respectively. $\tdd_{\mathrm{ref}}$ is the distance of the reference molecule from its initial position at time $\tau=0$, while $\tdd_{\mathrm{1st}}$ is the mean distance covered by all molecules belonging to the first shell at time $\tau$, since $\tau=0$.}
    \label{fig:scheme}
\end{figure}

To search for causal links underlying the complex dynamics of water, we used several CVs (illustrated in Figure~\ref{fig:scheme}) that directly probe the rotational and translational degrees of freedom in the local molecular environment of water~\cite{zanotti_experimental_2005}. For each trajectory in our dataset, we randomly selected one reference water molecule as the center of the local environment. We followed the dipole vector of this reference molecule and the weighted sums of its first and second shell water dipoles. 
These three CVs are denoted as $\bm\mu_{\mathrm{ref}}$, $\bm\mu_{\mathrm{1st}}$, and $\bm\mu_{\mathrm{2nd}}$, respectively. To investigate translational motion, we computed the distances traveled by molecules from an initial time $t_0$, both for the reference and for the water molecules in the first shell ($\tdd_{\mathrm{ref}}$ and $\tdd _{\mathrm{1st}}$). More details on CV construction are provided in \hyperref[section:Method]{Methods}. 

We separately analyzed room temperature and supercooled water simulations, in order to compare the dynamical couplings between pairs of CVs revealed by the IG measure.
Most of our results use the TIP4P/2005 water model~\cite{abascal_general_2005}, an empirical potential based on pairwise additive interactions. This water model has recently been shown to present critical-like fluctuations under supercooled conditions~\cite{debenedetti_second_2020}, between a high-density liquid (HDL) and a low-density liquid (LDL). The HDL liquid can be stabilized at 178 K and 200 MPa, and its dynamical properties can be converged on the microsecond timescale~\cite{debenedetti_second_2020}. To assess the fidelity of our results, we compare them against the state-of-the-art many-body polarizable (MB-pol)~\cite{palos_current_2024, zhu_mb-pol2023_2023} model both at room temperature and at supercooled conditions.

\begin{figure}[t!]
    \centering
    \includegraphics[width=0.8\linewidth]{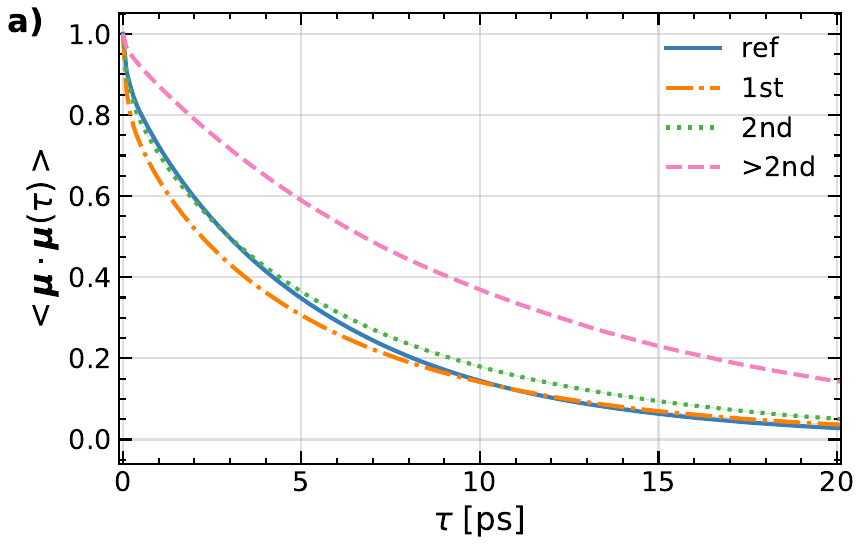}\\
    \includegraphics[width=0.8\linewidth]{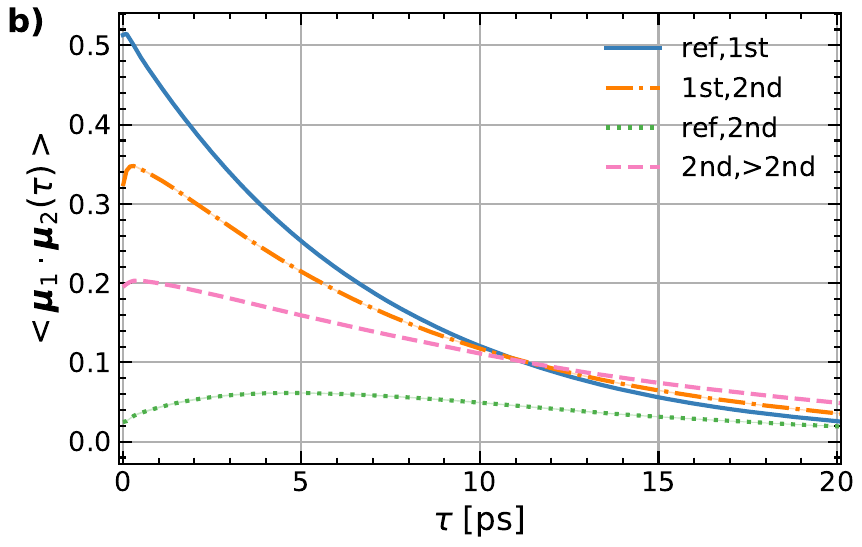}
    \caption{Autocorrelation and cross-correlation functions of the dipole moments studied in the system at room temperature, using the TIP4P/2005 water model.
    In the legend, the dipole of water molecules beyond the second solvation shell is denoted by ``$>$2nd''.
    The cross-correlations functions, in panel b, are rescaled by {\small$\sqrt{\langle\mrmu_1^2\rangle \cdot\langle\mrmu_2^2\rangle}$}.}
    \label{fig:roomtemp_correlations}
\end{figure}

\begin{figure*}[t!]
    \centering
    \includegraphics[width=0.32\linewidth]{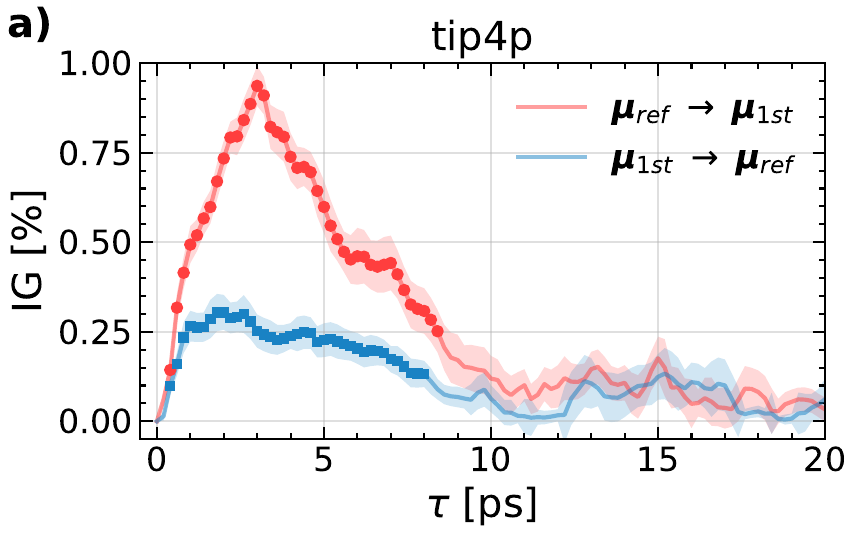}    
    \includegraphics[width=0.32\linewidth]{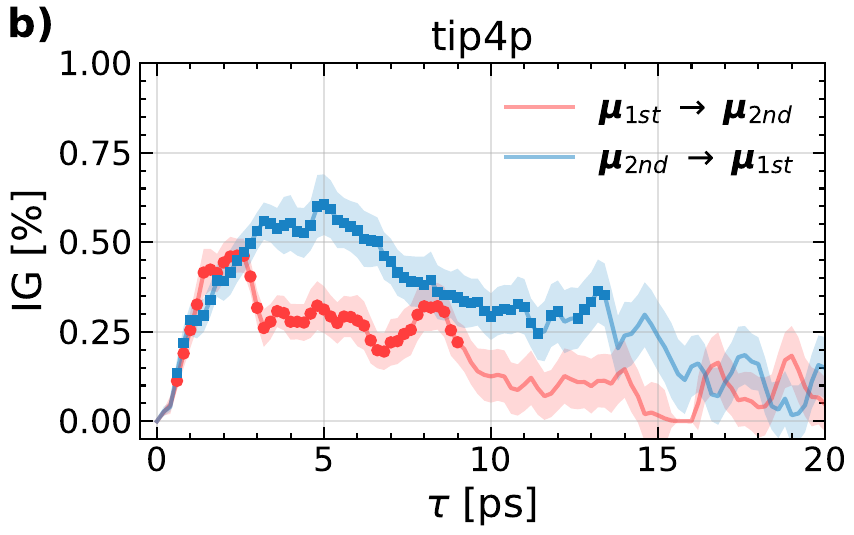}    
    \includegraphics[width=0.32\linewidth]{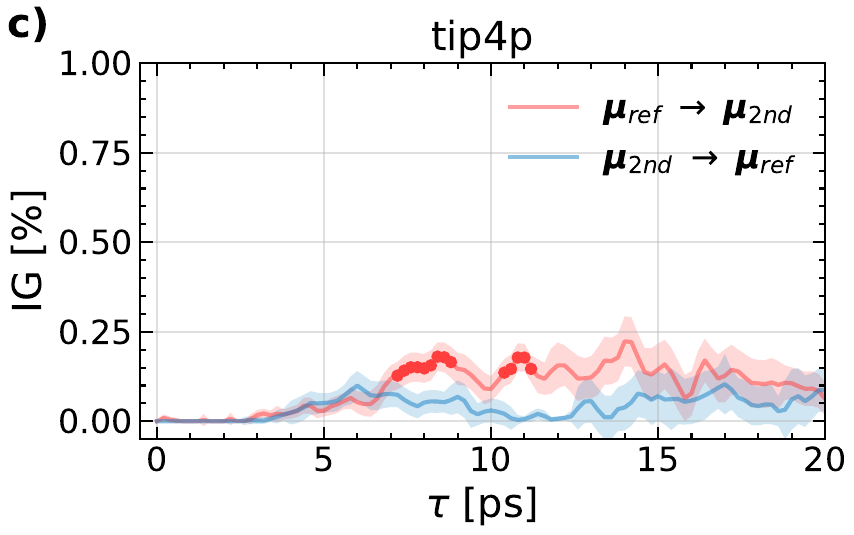}\\   
    \includegraphics[width=0.32\linewidth]{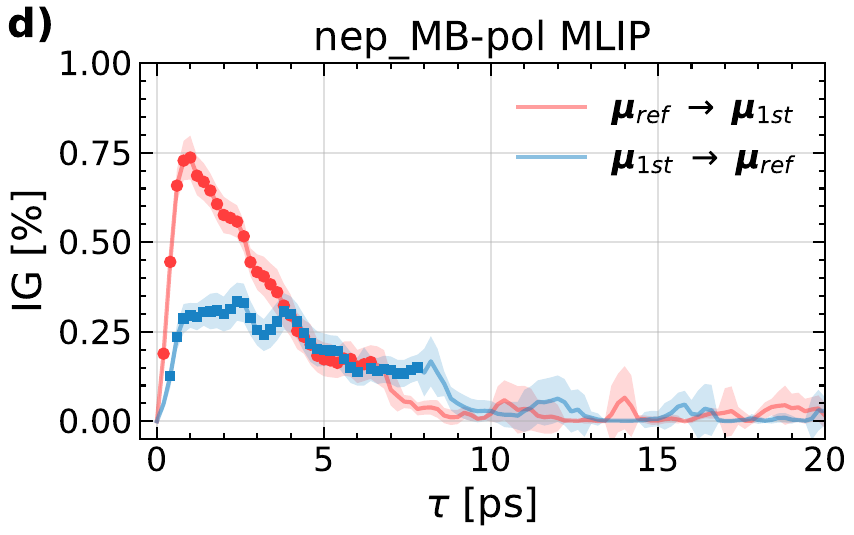}    
    \includegraphics[width=0.32\linewidth]{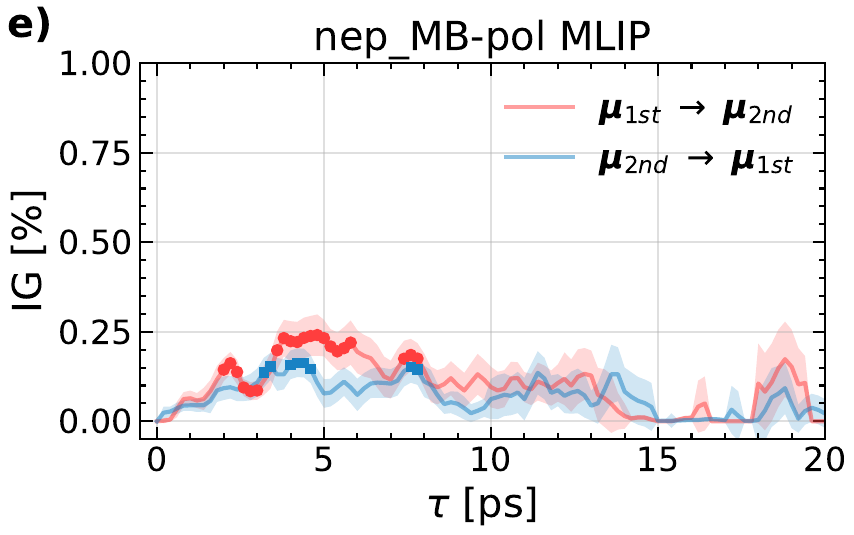}    
    \includegraphics[width=0.32\linewidth]{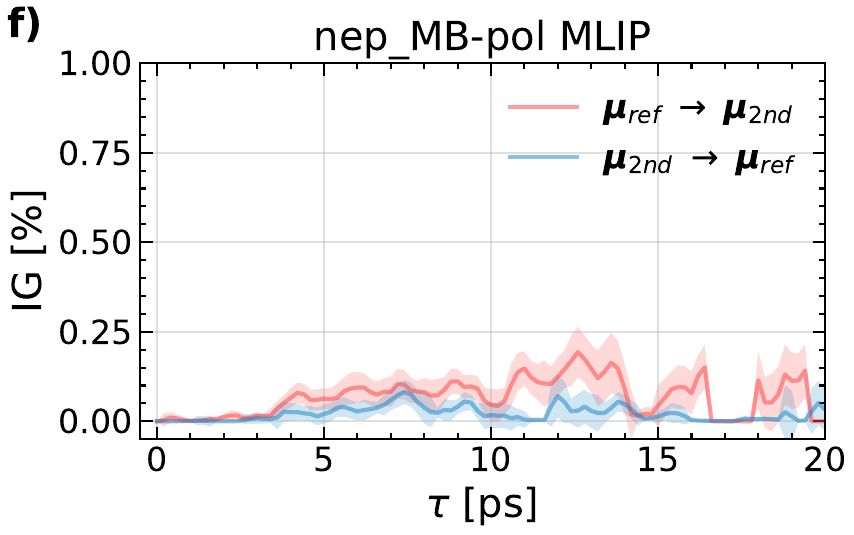}    
    \caption{Estimation of the Imbalance Gain at ambient conditions ($P=1$ atm, $T=300$ K). Each panel presents the results for a specific pair of variables in both directions. Panels a-c) show the IGs measure on the TIP4P/2005 force field dataset, while panels d-f) are referred to the MLIP~\cite{xu_nep-mb-pol_2025} trajectories. 
    }
    \label{fig:roomtemp_IG}
\end{figure*}

We first examined standard time-correlation functions. This allows us to establish the relevant dynamical timescales, but also to reveal the intrinsic limitations of these statistics.  Figure~\ref{fig:roomtemp_correlations} shows the autocorrelation functions at room conditions of $\bm\mu_{\mathrm{ref}}$, $\bm\mu_{\mathrm{1st}}$, $\bm\mu_{\mathrm{2nd}}$ and the dipole of all remaining molecules in the box, excluding the first two shells. The same curves in HDL conditions are presented in Supporting Information (SI) Figure S1.
The first three exhibit nearly identical exponential relaxation with a characteristic time of $\sim$5.5 ps, indicating dynamics dominated by single-molecule reorientation~\cite{moilanen_water_2008}. In contrast, the collective dipole relaxes more slowly ($10.51\pm0.02$ ps), reflecting cooperative motion at larger length scales~\cite{popov_mechanism_2016}. The cross-correlation functions in the bottom panel of Figure~\ref{fig:roomtemp_correlations} reveal a finite lag in the buildup of correlations between the reference dipole and the outer shells, indicating coupling between dipolar fluctuations across length scales. However, under detailed balance, these cross-correlation functions are invariant under exchange of the two variables of interest, and therefore cannot encode directional information or suggest causal links.

To identify asymmetric couplings that may underlie causal links, we computed the IG between pairwise combinations of $\mrmu_{\mathrm{ref}}$, $\mrmu_{\mathrm{1st}}$ and $\mrmu_{\mathrm{2nd}}$, presented in Figure~\ref{fig:roomtemp_IG} (top panels \textbf{a)-c)} for TIP4P/2005, bottom panels \textbf{d)-f)} for the MB-pol machine-learned interatomic potential (MLIP)~\cite{xu_nep-mb-pol_2025,xu_gpumd_2025}, both at room temperature conditions). We recall that the IG from $\mathbf{A}$ to $\mathbf{B}$ captures how much better $\mathbf{B}$ can be predicted when $\mathbf{A}$ is taken into account, thereby serving as a probe of causal influence in the spirit of Granger causality. This is quantified as the relative reduction in a rank-based measure of prediction error (see Imbalance Gain Estimation in \hyperref[section:Method]{Methods}) and reported as a percentage.
When the IG is non-zero, we infer the presence of a causal link between the corresponding CVs. The IG is evaluated as a function of the time lag $\tau$ between the ``present'' and the ``future'', with peaks identifying the characteristic delays at which the influence of the driver is most strongly manifested in the driven variable.
In Figure~\ref{fig:roomtemp_IG} and throughout the paper, shaded regions indicate the standard error of the mean computed 
over 50 IG estimates (see \hyperref[section:Method]{Methods}), and markers indicate values that are significantly non-zero ($p < 10^{-3}$, one-tailed t-test).

For both water models, the IG weakens in magnitude and develops an increasing lag by moving to outer shells. This indicates that dynamical couplings become both attenuated and delayed for CVs describing increasingly extended spatial environments, as well as for environments that are farther apart.
Both panels a) and d) reveal a clear directional asymmetry, which implies that $\mrmu_{\mathrm{ref}}$ is driving $\mrmu_{\mathrm{1st}}$.
We remark that the term ``driving'' is used here to highlight the most important coupling direction in a link that is in principle bidirectional.
The IG in both directions rises and decay over several picoseconds, consistently with hydrogen-bond network fluctuations occurring on comparable timescales~\cite{laage_molecular_2006,offei-danso_collective_2023, moilanen_water_2008, stirnemann_communication_2012, moilanen_ionwater_2009, ramasesha_ultrafast_2011}. 
Although the significant IG values in Fig.~\ref{fig:roomtemp_IG} are smaller than 1$\%$, we note that this scale is consistent with that of previous high-dimensional and real-world applications of the same measure~\cite{del_tatto_robust_2024}.

\begin{figure*}
    \centering
    \includegraphics[width=0.32\linewidth]{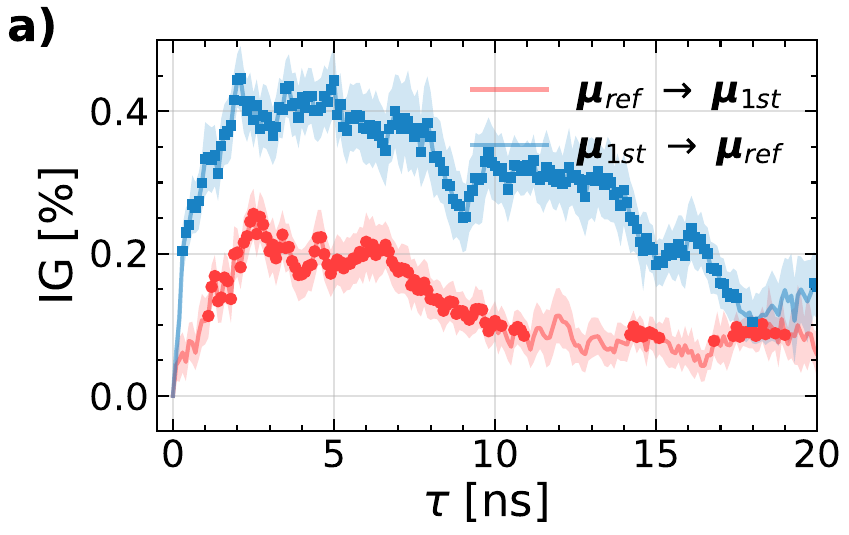}
    \includegraphics[width=0.32\linewidth]{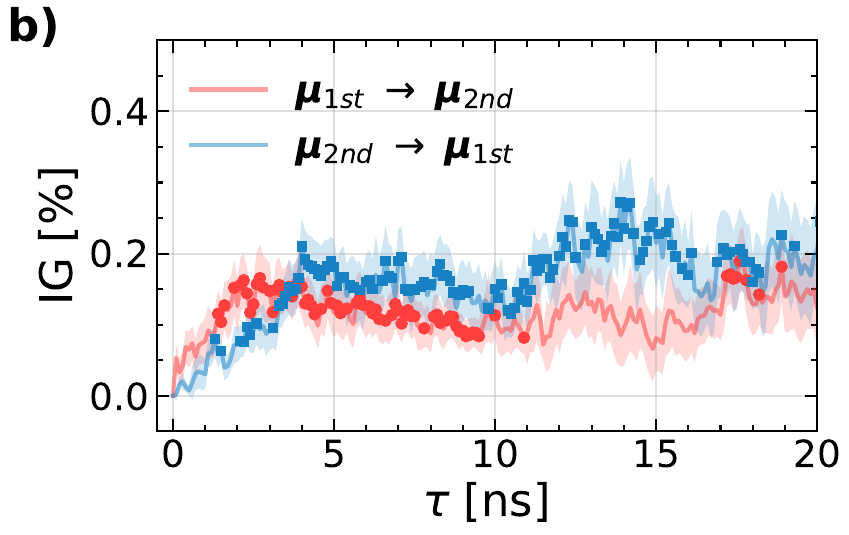}
    \includegraphics[width=0.32\linewidth]{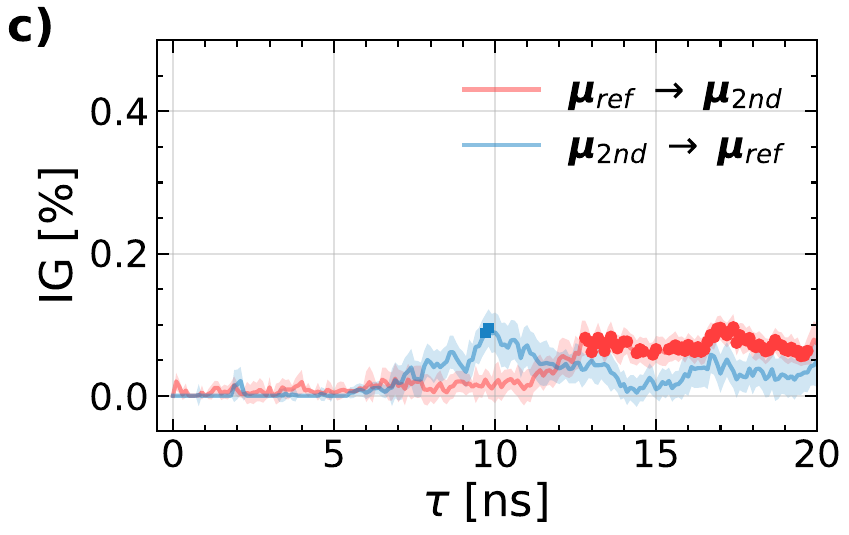}
    \caption{The estimation of Imbalance Gain between dipole moments for HDL condition ($T=178$ K, $P=200$ MPa) with the TIP4P/2005 force-field.} 
    \label{fig:HDL_IG}
\end{figure*}

In panel b), the coupling between $\mrmu_{\mathrm{1st}}$ and $\mrmu_{\mathrm{2nd}}$ appears perfectly bidirectional for $\tau \leq 2.5$ ps, followed by a crossover where $\mrmu_{\mathrm{2nd}}$ exerts a weak driving influence on $\mrmu_{\mathrm{1st}}$. Using the MLIP water model (panel e)), the inferred couplings links exhibit a weaker directional asymmetry.
Finally, panels c) and f) show almost negligible couplings between $\mrmu_{\mathrm{ref}}$ and $\mrmu_{\mathrm{2nd}}$, with a faint and delayed signal emerging (only for the TIP4P/2005 simulations) at $\sim$10 ps in direction $\mrmu_{\mathrm{ref}}\rightarrow\mrmu_{\mathrm{2nd}}$.

Having established the dynamical couplings at ambient conditions, we next examine their evolution in the supercooled regime. This region is central to ongoing debates surrounding a proposed liquid–liquid transition between high-density (HDL) and low-density (LDL) water, believed to underlie many anomalous properties. While experimental access is hindered by rapid ice nucleation, simulations provide evidence for critical-like HDL–LDL fluctuations and emerging dynamical heterogeneity~\cite{debenedetti_second_2020, sciortino_constraints_2025, xu_nep-mb-pol_2025, wang_direct_2026}. Specifically, for the HDL phase, since the onset of the diffusive regime occurs on nanosecond timescales, this regime offers a window into how slower fluctuations reshape directional dynamical couplings compared to room temperature conditions~\cite{malosso_dynamical_2025}.

We sampled water configurations harvested from long simulations of the TIP4P/2005~\cite{abascal_general_2005} model in a regime where the HDL phase is fully stabilized~\cite{debenedetti_second_2020}. 

Figure~\ref{fig:HDL_IG} displays the IG curves for the same dipole variables analyzed at room temperature conditions.
As in the case for room temperature water, we observe a progressive decrease in the IG magnitude and a dilation in the characteristic lag of the IG curves, going from pairs of CVs that represent spatially close environments to CVs that are ``farther apart''.

As expected, the associated timescales are shifted to the sub–10 ns regime. 
The main qualitative change emerges in the coupling between $\mrmu_{\mathrm{ref}}$ and $\mrmu_{\mathrm{1st}}$ (Figure~\ref{fig:HDL_IG}\textbf{a}), for which the 
first-shell dipole now acts as the driver of the reference dipole, with a clear inversion compared to ambient conditions. This inversion is confirmed also with the MB-pol water model analyzed from previous studies by
Sciortino \textit{et al.}~\cite{sciortino_constraints_2025} (see SI Figure S2).

A similar inversion is also observed in the LDL phase (see SI Figure S3). However, the highly reduced mobility in LDL water~\cite{malosso_dynamical_2025,  eltareb_evidence_2022, eltareb_isotope-substitution_2025, poole_dynamical_2011} prevents the extraction of independent initial conditions from a single trajectory 
(see SI section S4), making reliable IG estimates more challenging. Nonetheless, we observe similar trends in the LDL phase as we do for HDL. The qualitative difference between ambient and supercooled conditions hints at the possibility of other collective variables that may play an important role in affecting the cooperative fluctuations of the hydrogen-bond network.

A hallmark of supercooled liquid dynamics is the transient confinement of particles on intermediate timescales, commonly referred to as ``caging''~\cite{SCIORTINO2000307,matubayasi2019}. This phenomenon arises because particle motion requires the restructuring or partial dissolution of the surrounding solvation shell, thereby enabling escape from the cage. Such processes are not directly captured by the dipole collective variables, which primarily encode rotational degrees of freedom. Therefore, we extend our analysis to the mutual relationships between the dipolar and translational collective variables $\tdd_{\mathrm{ref}}$ and $\tdd_{\mathrm{1st}}$.

\begin{figure}[ht!]
    \centering
    \includegraphics[width=\linewidth]{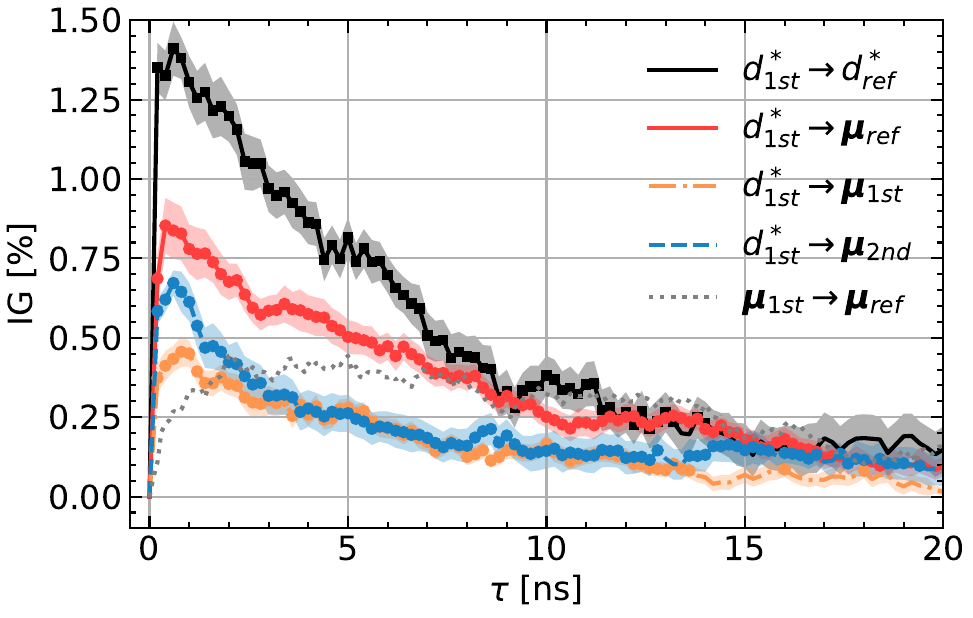}
    \caption{Selection of the IG curves in HDL conditions for which we observed strong couplings that are not present in the ambient conditions. The curve $\mrmu_{\mathrm{1st}} \rightarrow \mrmu_{\mathrm{2nd}}$ is also shown for comparison with the results of Figure \ref{fig:HDL_IG}.}
    \label{fig:Caging-effect}
\end{figure}

Figure~\ref{fig:Caging-effect} shows the IG curves obtained for HDL water, focusing on the most relevant cross-couplings between dipolar and translational CVs. Remarkably, we observe that $\tdd_{\mathrm{1st}}$, which essentially encodes the translational motions of the first solvation shell, emerges as the leading driver of all the dipolar variables (see SI Figures S5 to S7). All these couplings rise and decay within a time scale of 10 ns. Within this context, the directional couplings associated with $\tdd_{\mathrm{1st}}$ acting as the driver are consistent with the dynamics of the crossover from caging to the diffusive regime observed in HDL~\cite{malosso_dynamical_2025}. Moving beyond the first shell, we also observe that $\tdd_{\mathrm{2nd}}$ appears as the main driver of $\tdd_{\mathrm{1st}}$, with a coupling which persists up to
$\sim$40 ns (see SI Figure S8). Importantly, while these results highlight a dominant directional influence of translational dynamics on dipolar variables, they do not preclude reciprocal couplings: orientational and translational motions can in fact be coupled in a bidirectional manner, as illustrated in SI Figure S5 \textbf{c)}-\textbf{g)}. These directional couplings—across translational length scales and between translational and orientational motion—reflect a facilitation mechanism which is enhanced under supercooled conditions.

\begin{figure}[t!]
    \centering
    \includegraphics[width=\linewidth]{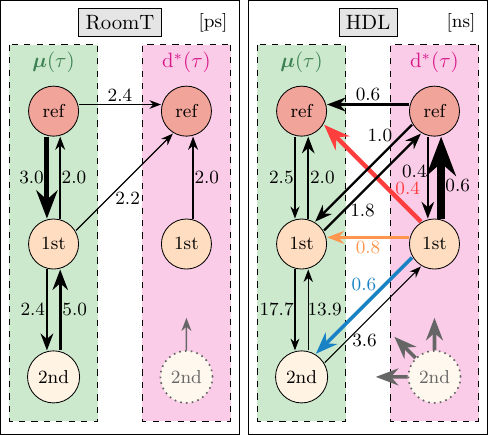}
    \vspace{-1.5em}
    \caption{Summary diagram of the pairwise IG estimations made in this study for the TIP4P/2005 force-field~\cite{abascal_general_2005}. Arrows between two nodes are drawn only when significantly non-zero values are observed in the corresponding IG curves. Arrow thicknesses are proportional to the maxima of the IG curves. Labels report the time delays corresponding to these maxima, in ps for the ambient condition and in ns for the HDL regime. Colored arrows in the HDL diagram correspond to the curves shown in Figure \ref{fig:Caging-effect}.}
    \label{fig:diagram}
\end{figure}

\correc{The same analysis was carried out for the room-temperature system, revealing weak but meaningful directional couplings from the rotational (dipolar) CVs to the translational CVs (see SI Figure. S6). Figure~\ref{fig:diagram} summarizes all pairwise couplings measured by the IG for the collective variables considered in this work, under both ambient and HDL conditions. The corresponding curves are presented in Figures 
\ref{fig:roomtemp_IG}, \ref{fig:HDL_IG},
S5, and S6. Nodes represent specific CVs, while arrows denote pairwise couplings inferred from the IG curves. The thickness of each arrow is proportional to the maximum value of the IG across the tested lags, and the number reported next to each arrow indicates the lag (time) at which this maximum is observed. At room temperature, the dynamics are dominated by couplings within the translational and dipolar sectors, with the presence of a weak but detectable directional coupling from dipolar to translational variables. In contrast, under HDL conditions, translational degrees of freedom emerge as the primary drivers of the dynamics, with most couplings directed from translational to dipolar CVs.}

\correc{These differences in the coupling topology also suggest distinct microscopic mechanisms governing the dynamics in the two thermodynamic regimes. 
At room temperature, 
the orientation of first-shell molecules is influenced not only by the reference molecule but also by the surrounding second hydration shell, since hydrogen bonds extending beyond the first shell continuously modify the local orientational constraints. This interpretation is consistent with the signal propagation observed in Figure \ref{fig:roomtemp_IG}(a)-(c), where the inferred information transfer extends sequentially through neighboring hydration shells. Rather than being mediated solely by the reference molecule, orientational perturbations appear to propagate through the transient hydrogen-bond network together with local density fluctuations. This picture is also consistent with our previous proposal that angular jumps occur as cooperative bursts involving correlated orientational rearrangements and density fluctuations \cite{offei-danso_collective_2023}. 
Under HDL conditions, by contrast, the predominance of translational-to-dipolar couplings indicates that orientational relaxation becomes increasingly constrained by the slowly evolving local structure, consistent with the emergence of caging effects in the supercooled liquid (see SI Figure S10 comparing the MSD in room temperature and supercooled HDL water).}

\section{Discussion and Conclusion}

The dynamics of supercooled liquids are marked by particle caging, facilitation, and heterogeneous motion, with clusters of fast and slow particles emerging alongside growing correlation lengths. Yet, despite decades of work, their microscopic origin remains elusive. Most signatures of heterogeneity are inferred indirectly—often from simulations—and there is still no direct way to probe how microscopic motions influence one another. In this context, by importing ideas from causal inference, we identify directional couplings between molecular degrees of freedom in liquid water—from ambient to deeply supercooled conditions—that remain invisible to standard correlation-based approaches.

Against this backdrop, Figure~\ref{fig:diagram} distills the central result. \correc{At ambient conditions, the inferred directional coupling between translational and rotational dynamics is comparatively weak but nevertheless statistically significant. Specifically, we identify a weak directional influence from the rotational (dipolar) collective variables to the translational collective variables. This observation is consistent with the established picture that translational and rotational motions in liquid water is  coupled, as reflected in the approximate validity of the Debye-Stokes-Einstein relation \cite{mazza_connection_2007}, quasi-elastic neutron scattering measurements \cite{blanckenhagen_intermolecular_1972, qvist_structural_2011}, and the microscopic jump mechanism proposed by Laage and Hynes \cite{laage_molecular_2006}, which links local translational rearrangements within the first hydration shell to molecular reorientation. 
In particular, the observed rotational-to-translational coupling is consistent with the observation that large reorientations tend to occur in, and to generate, low-density regions of the network~\cite{offei-danso_collective_2023}: if transient caging is tied to local density fluctuations, reorientational events that create low-density defects could in turn facilitate the local translational rearrangements associated with cage relaxation.
Our analysis complements these previous studies by quantifying the directional asymmetry of the coupling directly from the molecular dynamics.}

\correc{
Using rotational variables we find that at ambient conditions the dominant couplings are directed from the central molecule to the first shell ($\tau \sim 1$--$8$~ps); that the first and second shells are strongly and almost symmetrically linked ($\tau \sim 1$--$15$~ps); and that a weak, unidirectional influence from the central molecule to the second shell emerges only at longer lags ($\tau \sim 10$~ps). 
This hierarchy is what one would expect in a scenario in which a reorientation of the central molecule triggers a cascade of correlated reorientations propagating outward through the hydrogen-bond network: the central molecule directly reorients its first-shell neighbours and reaches the second shell only indirectly, through the first.
The resulting decay of directed influence with distance is consistent with the cooperative reorientation mechanism of Ref.~\cite{offei-danso_collective_2023}, in which angular jumps occur in propagating ``waves'' and a reorientation enhances subsequent jumps in its neighbourhood.
To support this interpretation, we note that the characteristic lag times we observe ($\tau \sim 1$--$15$~ps) lie on the collective, many-molecule timescale of Ref.~\cite{offei-danso_collective_2023}, rather than on the sub-picosecond timescale of an individual jump.
}

\correc{Upon supercooling into the HDL regime, the pattern of directional coupling changes qualitatively. The weak rotational-to-translational influence observed at ambient conditions is replaced by a pronounced directional influence of translational motion on rotational dynamics, while the reverse direction becomes comparatively much weaker, as evidenced by the dominant $\tdd \rightarrow \mrmu $ links.} 
Crucially, the absence of a link is itself informative—within the causal framework, it rules out direct influence even in the presence of hidden common drivers~\cite{spirtes_causation_2001}. 
\correc{A natural molecular interpretation of this reversal is that, upon supercooling, reorientation becomes limited by the local cage: unlike at ambient conditions, a molecule can no longer reorient through a simple hydrogen-bond exchange, but has to wait for the  structural rearrangement of its neighbors. The reorientation is therefore controlled by the translational relaxation of the environment, rather than the other way around.}
This asymmetry provides a microscopic mechanism for dynamical heterogeneity: highly mobile translational motions drive the relaxation of neighboring, less mobile dipoles. In this way, our results provide a more nuanced understanding of the long-standing picture of translational–rotational decoupling in supercooled liquids~\cite{mazza_connection_2007, dueby_decoupling_2019,nagaraj_structural_2026}, showing that decoupling is not merely a breakdown of proportionality, but the emergence of directionality: rotational motion becomes driven by translations, while the reverse pathway fades. At the same time, strong couplings between translational degrees of freedom persist across neighboring molecules, supporting a facilitation mechanism propagated through translational motion.
This directional asymmetry carries important physical consequences. Under supercooled conditions, perturbations to rotational degrees of freedom should have little measurable impact on translational motion over nanosecond timescales, whereas perturbations of translation should propagate to rotations in the same time interval. Interventions on translation would therefore generate a stronger arrow of time, as their effects cascade across degrees of freedom.

Viewed through this lens, an important question is how robust this directional picture is to the underlying molecular description. Strikingly, the qualitative structure of the inferred causal graph remains unchanged across two markedly different water models—the rigid, non-polarizable TIP4P/2005 potential and the many-body polarizable MB-pol model. In both cases, causal influence propagates outward across hydration shells, and supercooling gives rise to the same translation-to-rotation asymmetry. This robustness suggests that the observed directionality reflects an underlying physical feature. At the same time, quantitative differences are evident. MB-pol consistently yields weaker IG signals—particularly for dipolar couplings—indicating that the strength of causal links is sensitive to how intermolecular interactions and fluctuations are represented. The explicit treatment of many-body polarization in MB-pol likely introduces greater configurational variability, thereby reducing the predictive leverage between degrees of freedom compared to the more constrained TIP4P/2005 model. In this sense, causal metrics not only uncover directional couplings, but also provide a sensitive lens on how different interaction models encode molecular fluctuations.

\correc{
Related directional couplings between different classes of intermolecular motion have already been explored in ambient liquid water using multidimensional nonlinear spectroscopies. In particular, Yagasaki and Saito~\cite{yagasaki_molecular_2009} used molecular-dynamics simulations of 2D Raman and third-order IR responses to show that excitation of low-frequency translational modes leads to a strongly anharmonic response and pronounced coupling between translational and librational motions, and that translational motion has a significant influence on the fast frequency modulation of librations. These studies demonstrate that out-of-equilibrium multidimensional spectroscopy can indeed be used to infer time-delayed, directional influences between distinct modes and thereby to reveal an explicit arrow of time in the dynamical response. 
However, the directional couplings they report develop and decay within a few hundred femtoseconds, faster than the hydrogen-bond network can rearrange.
The directed couplings quantified here through the IG, in contrast, are evaluated on the ps timescale, at the onset of structural relaxation, where the relevant rotational motion is no longer the inertial libration but the diffusive reorientation of the molecules. 
A direct comparison between the two analyses is therefore not straightforward; nonetheless, at the shortest lags we access ($\tau \sim 0.5$~ps), which lie closest to the regime probed by multidimensional spectroscopy, we detect a weak directional coupling oriented from translational to rotational degrees of freedom at ambient conditions (Fig.~S6, panels b and c), consistent with the findings of Yagasaki and Saito.
}

\correc{An important open question, which we aim to address in future work, is 
to what extent the equilibrium directional couplings we infer in the HDL phase of supercooled water are predictive of actual causal links, revealed by suitable interventional experiments.
This could be probed experimentally using pump–probe spectroscopy: a tailored pump excites either rotational or translational modes, while a delayed probe monitors the induced response in the complementary degree of freedom, for example by extending the multidimensional spectroscopic approaches of Ref.~\cite{yagasaki_molecular_2009} to supercooled conditions. 
More broadly, these results open a route toward experimentally accessing directional couplings in molecular liquids. At the same time, an important open question concerns the role of potential confounding variables in shaping the inferred causal structure from equilibrium molecular dynamics. Recent advances in multivariate causal discovery~\cite{allione_linear_2025} provide a promising path forward, although extending these approaches to deeply supercooled systems remains computationally demanding.}
 
\label{section:Method}\section{Methods}

\subsection*{Molecular Dynamics Simulations}

The studied system consists of a cubic box containing 300 water molecules, integrated with a 0.5 fs timestep.
Molecular dynamics simulations using the classical empirical TIP4P/2005~\cite{abascal_general_2005} water model were performed with the GROMACS software package~\cite{abraham_gromacs_2015}. All simulations were carried out in the NPT ensemble, using a Nosé–Hoover chain thermostat~\cite{nose_unified_1984, hoover_canonical_1985,  martyna_nosehoover_1992} to control temperature and a Parrinello–Rahman barostat~\cite{parrinello_polymorphic_1981} to maintain constant pressure. Two thermodynamic conditions were considered for these simulations. The ambient-conditions' simulations were performed at 300 K and 1 atm, while the HDL simulations were conducted at 178 K and 200 MPa, which have previously been shown to stabilize this phase~\cite{debenedetti_second_2020}.

The machine-learning interatomic potential (MLIP) simulations were performed under the same thermodynamic conditions as the classical room-temperature simulations. These simulations were carried out using the GPUMD code~\cite{xu_gpumd_2025} together with the NEP force field, trained on MB-pol reference simulations by Xu \textit{et al.}~\cite{xu_nep-mb-pol_2025} For the equivalent analysis in the HDL phase, the trajectory has been obtained from a previous study by Sciortino \textit{et al.} at 188K and 120MPa~\cite{sciortino_constraints_2025}.

\subsection*{Water Collective Variables (CVs)}

To define the boundaries of the solvation shells used for the construction of the CVs, we used the oxygen-oxygen radial distribution function (RDF). Specifically, we chose the first two local minima as limits for the first and second shells: $d^\circ_{1} = 3.3$~\AA~and $d^\circ_{2} = 5.7$~\AA. In order to ensure that the CVs were smooth and continuous, each oxygen atom was dressed with a  smooth Fermi-like switching function defined as the following:
{\small \begin{align}
    w_{\text{1st}}(i) = & \frac{1}{1+ e^{\frac{d_{i}-d^\circ_{1}}{s}}}\,, \\
\begin{split}
    w_{\text{2nd}}(i) = &\frac{1}{1 + e^{\frac{d^\circ_{1}-d_i}{s} }} + \frac{1}{1 + e^{\frac{d_{i}-d^\circ_{2}}{s}}} - 1 \,,
\end{split} 
\end{align}}where $s$ is a smoothness parameter set to 0.1~\AA~ and $d_i$ is the distance between the reference atom and the oxygen belonging to the $i$-th water molecule in the box. For the HDL phase we used the same shell boundaries, as the RDF minima are located at the same values. 
For the reference molecule, $w_{\text{ref}}$ can be trivially defined as 1 when $i$ identifies the reference oxygen, and zero otherwise.
The dipoles of the shells are defined as weighted sums:

\begin{equation}
    \mrmu _\alpha = \sum_i w_\alpha(i) \mrmu_i\,,
\end{equation}
where $\mrmu_i$ is the dipole moment of each water molecule $i$, and $\alpha \in \{\text{ref},\,\text{1st},\,\text{2nd} \}$.

CVs that describe translational motion are defined as weighted averages of the distances traveled by the oxygen atoms since $\tau=0$. For convenience, these CVs were further centered by setting their mean values at zero for each value of $\tau$:
\begin{equation}
    \tdd _\alpha = \frac{\sum_i w_\alpha(i) \tdd_i}{\sum_i w_\alpha(i)} - \bigg< \frac{\sum_i w_\alpha(i) \tdd _i}{\sum_i w_\alpha(i)} \bigg>\,,
\end{equation}
where $\tdd_i$ is the distance of the oxygen of molecule $i$ from its initial position: $|| \bm{r}_i(\tau)-\bm{r}_i(0)||$.
The mean value $\langle\cdot\rangle$ was calculated across independent trajectories (see section Dataset Creation) at each time $\tau$.
It is important to note that setting the mean value of these variables to 0 does not affect the IG estimation since the distance ranks used in its definition are invariant under translations in the feature space (see section Imbalance Gain Estimation) .

\subsection*{Imbalance Gain Estimation}

Information Imbalance (II) and Imbalance Gain (IG) were introduced in Refs.~\cite{glielmo_ranking_2022,del_tatto_robust_2024}. The codes used to estimate II and IG are available in the DADApy python package~\cite{glielmo_dadapy_2022}. In the following, these methods are briefly summarized and adapted to the context of the present study.

The II allows for a quantitative comparison of different distance measures, by examining how local neighborhood relationships change from one distance to another.
Given a dataset of $N$ points independently drawn from an underlying probability distribution, we consider two distances $\metricd^A$ and $\metricd^B$, and denote by $r_{ij}^A$ ($r_{ij}^B$) the distance rank (neighbor order) of point $j$ with respect to $i$ in the corresponding distance space.
For example, $r_{ij}^A = 2$ if $j$ is the second nearest neighbor of $i$ among all points in the dataset, according to distance $\metricd^A$.
The II from $\metricd^A$ to $\metricd^B$ is defined as:
\begin{equation}\label{eq:II}
\begin{split}
    \Delta(\metricd^A \rightarrow \metricd^B) &= \frac{2}{N} \left< r^B | r^A \leq k \right> \\
    & =   \frac{2}{N^2k} \sum_{r^A_{ij} \leq k} r^B_{ij}\,,
\end{split}
\end{equation}
where $k$ is a hyperparameter of the approach, denoting the number of nearest neighbors taken into account. In this work, we used $N=2000$ and $k = 20$ in all our analysis. 
The prefactor in Eq.~(\ref{eq:II}) allows us to statistically confine the II between 0 and 1; a value close to 0 denotes that neighborhood relationships identified in $\metricd^A$ are well reproduced in $\metricd^B$, while a value close to 1 denotes that nearest neighbors in space $\metricd^A$ have randomly distributed ranks according to $\metricd^B$.

The IG builds on the II to implement a Granger-like causality test, using distance spaces constructed over two CVs: $\mathbf{A}(t)$ and $\mathbf{B}(t)$.
Pairwise distances are computed between independent replicas of the same dynamics, which can also be obtained from uncorrelated samples of a single stationary trajectory, treated as independent initial conditions ($\tau=0$).
Given a target distance space built on $\mathbf{B}$ at time $t=\tau$, the IG measures the ``information gain'' obtained when the target is described using distances defined on both $\mathbf{A}$ and $\mathbf{B}$ at time $t=0$, compared with distances only including $\mathbf{B}(0)$.
Using the II, this can be translated into the inequality
{\footnotesize \begin{equation}\label{eq:II_ineq}
    \min_\alpha\Delta \left(\metricd^{\alpha \mathbf{A}(0), \mathbf{B}(0)} \rightarrow \metricd^{\mathbf{B}(\tau)} \right) < \Delta \left(\metricd^{\mathbf{B}(0)} \rightarrow \metricd^{\mathbf{B}(\tau)} \right)\,,
\end{equation}}%
where $\metricd^{\mathbf{B}(0)}$ and $\metricd^{\mathbf{B}(\tau)}$ are Euclidean distances built over $\mathbf{B}(0)$ and $\mathbf{B}(\tau)$, while $\metricd^{\alpha \mathbf{A}(0),\mathbf{B}(0)}$ denotes the two following ``mixed'' Euclidean distance between points $i$ and $j$:
{\footnotesize \begin{equation} 
    \metricd^{\alpha \mathbf{A}(0), \mathbf{B}(0)}_{ij} = \big[\alpha^2\|\mathbf{A}_i(0) - \mathbf{A}_j(0) \|^2 + \| \mathbf{B}_i(0) - \mathbf{B}_j(0)\|^2\big]^{\frac{1}{2}}\,.
\end{equation}}
\noindent The weight $\alpha > 0$ allows identifying the optimal combination of $\mathbf{A}(0)$ and $\mathbf{B}(0)$ to predict the target distance space, and plays the role of an optimization parameter.

Calling $\Delta(\alpha)$ and $\Delta(\alpha=0)$ the (LHS) and (RHS) of inequality (\ref{eq:II_ineq}), the IG is defined as
\begin{equation}\label{eq:IG_def}
    \text{IG}_{\mathbf{A}\rightarrow \mathbf{B}} (\tau) \doteq \frac{\Delta(\alpha=0) - \min_\alpha\Delta(\alpha)}{\Delta(\alpha=0)}\,.
\end{equation}
Inequality (\ref{eq:II_ineq}) is equivalent to $\text{IG}_{\mathbf{A}\rightarrow \mathbf{B}} > 0$.
\correc{We note that the IG does not depend on the choice of units (equivalently, on the overall fluctuation amplitude) of the CVs $\mathbf{A}$ and $\mathbf{B}$. 
This is because the II, and hence the IG, depends on the distance spaces only through their rank orderings, which are preserved under any uniform rescaling of a variable. 
The ranks of $\metricd^{\mathbf{B}(0)}$ and $\metricd^{\mathbf{B}(\tau)}$ are invariant under $\mathbf{B}\mapsto b\,\mathbf{B}$ for any constant $b$, while the ranks of the mixed space $\metricd^{\alpha\mathbf{A}(0),\mathbf{B}(0)}$ are preserved under the joint rescaling $\mathbf{A}\mapsto a\,\mathbf{A}$, $\mathbf{B}\mapsto b\,\mathbf{B}$ provided the weight is also scaled as $\alpha\mapsto\frac{b}{a}\,\alpha$.
In other words, redefining the units of $\mathbf{A}$ and $\mathbf{B}$ merely shifts the location of the optimal weight $\alpha$, leaving $\min_\alpha\Delta(\alpha)$ (and therefore the IG of Eq.~\ref{eq:IG_def}) unchanged.}

In practice, the minimization of the II appearing in Eq.~(\ref{eq:IG_def}) was carried out in this work by replacing $\Delta(\alpha)$ with
\begin{equation}
    \Delta(\beta) = \Delta\left(\metricd^{(1-\beta) \mathbf{A}(0),\, \beta \mathbf{B}(0)} \rightarrow \metricd^{\mathbf{B}(\tau)} \right)\,,
\end{equation}
minimized for $\beta \in [0,1]$ over a grid of
200 evenly spaced values. The two minimization approaches are equivalent, as they only differ for the definition of the optimization range ($\alpha\in[0,+\infty),\, \beta\in\left[0,1\right]$).


In the field of time-series analysis, using short trajectory segments rather than individual time points is a well-established practice, grounded in embedding theory~\cite{takens_detecting_1981} and already explored in the context of molecular dynamics~\cite{hegger_how_2007}. More formally, in Ref.~\cite{del_tatto_robust_2024} we showed that distance functions built over time-delay embeddings of $\mathbf{A}(t)$ and $\mathbf{B}(t)$, in the form 
\begin{equation}\label{eq:time-delay-embeddings}
    \big(\mathbf{A}(t),\, \mathbf{A}(t-\tau_e),\, ...,\, \mathbf{A}(t-(E-1)\tau_e)\big)
\end{equation}
(similarly for $\mathbf{B}(t)$), reduce false positives and lead to a more reliable identification of non-causal relationships.

In our analysis, we set $E=50$ and $\tau_e=10$ fs for the room conditions analysis, and $E=10$, $\tau_e=10$ ps for the HDL case. In both cases, $\tau_e$ corresponds to the sampling time of the trajectories.

Throughout this paper the IG was reported as a percentage, reflecting its definition as a relative difference (Eq.~(\ref{eq:IG_def})). In a dataset of 100 trajectories, an IG$(\tau) = 1\%$ from $\mathbf{A}$ to $\mathbf{B}$ indicates that including $\mathbf{A}$ in the prediction criterion improves the proximity to a reference realization by one neighbor order, on average, compared to when $\mathbf{A}$ is excluded ($\alpha=0$).

\correc{Unlike quantities such as correlation coefficients, IG does not admit a universal scale on which one can define \emph{small} or \emph{large} values. Instead, its characteristic magnitude depends on the dimensionality and complexity of the underlying dynamical system. 
Previous applications of the same methodology illustrate this trend: values approaching 80\% are observed for coupled Rössler dynamical systems\cite{del_tatto_robust_2024}, approximately 7\%  for two-dimensional Langevin dynamics \cite{del_tatto_towards_2025}, 3–6\% for atomistic molecular simulations of an amino acid in water \cite{del_tatto_towards_2025} and 1-2\%, in a recent work\cite{banerjee_investigating_2026}, between principal components of a folded protein dynamics. As the number of interacting degrees of freedom increases, the directional influence becomes distributed across an increasing number of coupled degrees of freedom, reducing the pairwise IG associated with any single descriptor.}

\correc{The physically meaningful outcome of an IG analysis are three: first, and most importantly, the absence of a coupling, which is inferred by values of the IG which are statistically consistent with zero. Second,  the relative changes in IG between  different thermodynamic condition, and, third, the difference in IG values  between different pairs of observables in the same conditions.}

\correc{In the paper of Del Tatto \textit{et al} in 2024\cite{del_tatto_robust_2024}, a synthetic system with a complexity comparable to the one of this work has been tested, the couple of two variable in a bath composed of two unidirectionally coupled Lorenz 96 systems, with 40 variables each. The measure was also performed using time-delayed embeddings and the final signal, when clearly unidirectional as expected, had an intensity of 3\%, see Figure \textbf{3T} in the cited manuscript.}

In the SI (section S7), we show that an information-theoretic reformulation of this measure allows to express the IG in information bits.

\subsection*{Dataset Creation}

For the room-temperature system, 50 independent trajectories of 200 ns each were generated. From each trajectory, 2000 evenly spaced segments of 50 ps were extracted, resulting in consecutive chunks with a separation of 100 ps between their initial frames.
Each set of 2000 segments extracted in this way formed an independent dataset used for the IG estimation. However, our measurements of autocorrelation and IG under room conditions (Figures \ref{fig:roomtemp_correlations} and \ref{fig:roomtemp_IG}) indicated that a trajectory of 100 ns with 50 ps between initial frames
was already sufficient for convergence (the signal is already null after $\sim$20 ps). The error bars were estimated as the standard errors of the mean over the 50 independent estimations of the IG, for each value of $\tau$. 
For the MLIP case, due to the computational cost of these simulations, the length of the 50 main trajectories was limited to 100 ns. The datasets used for the IG estimation were generated in the same way (2000 evenly spaced segments of 50 ps), resulting in consecutive segments with a separation of 50 ps between their initial frames.

Due to the significantly slower dynamics of supercooled water, generating multiple independent trajectories of comparable length under HDL/LDL conditions would be computationally prohibitive. We therefore adopted an approximation by considering the CV trajectories extracted from different local environments in the same system
as sufficiently independent of each other. Instead of conducting multiple independent simulations, we generated a single trajectory of 10 $\mu$s and randomly selected different central molecules to construct 50 equivalent datasets of CV trajectories. 
The resulting datasets are again composed of 2000 trajectory segments of 50 ns, evenly spaced every 5 ns, and partially overlapping with each other.
Unlike room-temperature analysis, the 50 datasets are not strictly independent, and the CV samples within each dataset may exhibit residual time correlations.
In the SI (See section S4) we motivate why these dependencies do not affect the results in HDL conditions.
\correc{Additionally, due to the weak signal associated with the rotational-to-translational terms in the initial room conditions IG estimation, we generated a dedicated dataset specifically to resolve this signal. This new dataset consists of 5000 trajectories of 20 ps, each repeated 100 times to improve the statistical resolution. It was obtained by combining the initial 50 independent trajectories used for the room-temperature simulations with the reference-molecule selection strategy employed for the HDL system.}

\begin{acknowledgement}

A.H., L.H., and D.B.  acknowledge the funding received by the European Research Council (ERC) under the European Union’s Horizon 2020 research and innovation program (Grant No. 101043272 - HyBOP). The views and opinions expressed are
those of the authors only and do not necessarily reflect those of the European Union or the European Research Council Executive Agency. Neither the European Union nor the granting authority can be held responsible for them. A.H., L.H., and D.B. also acknowledge CINECA for their resources on the cluster Leonardo (from the Convenzione triennale ICTP Anni 2024-2026 project) where all simulations and post processes were carried. L.H. thanks Florian Pabst and Francesco Paesani for sharing the trajectory of HDL in the supercooled regime for the MB-pol water model. A.H.  acknowledges the use of ChatGPT for improving the text and readability of the manuscript. All coauthors acknowledge Edward Danquah Donkor for performing some preliminary analysis.

\end{acknowledgement}

\begin{suppinfo}

The supplementary material includes supporting figures for cross-correlation and IG estimates, results for the LDL phase of supercooled water, and simulation details.

The authors declare no conflict of interest.

\correc{Inputs and codes are available on Git-Hub: \href{https://github.com/HUETLeon/Code-Causality-in-Liquid-Water-as-a-Hallmark-of-Emergent-Glassy-Dynamics.git}{https://github.com/HUETLeon/Code-Causality-in-Liquid-Water-as-a-Hallmark-of-Emergent-Glassy-Dynamics.git}. Datasets of CVs and imbalance estimations are separately available on a Zenodo repository: \href{https://doi.org/10.5281/zenodo.20157241}{https://doi.org/10.5281/zenodo.20157241}, and can be used to reproduce the steps of our work from II estimation to figure's creation. Large trajectories are available on demand to the author.}

L.H. performed the simulations, produced the figures, and participate in writing the manuscript. V.D.T. and D.B. developed the theoretical framework and contributed to the analysis and interpretation of the results; V.D.T. also contributed to writing the manuscript, while D.B. provided critical revisions. A.L. and A.A.H. conceived and supervised the research, contributed to the interpretation of the results, and participated in writing the manuscript. All authors discussed the results and approved the final version of the manuscript.

\end{suppinfo}

\bibliography{references}

~

\newpage

~

\newpage

\newpage

\includepdf[pages=-]{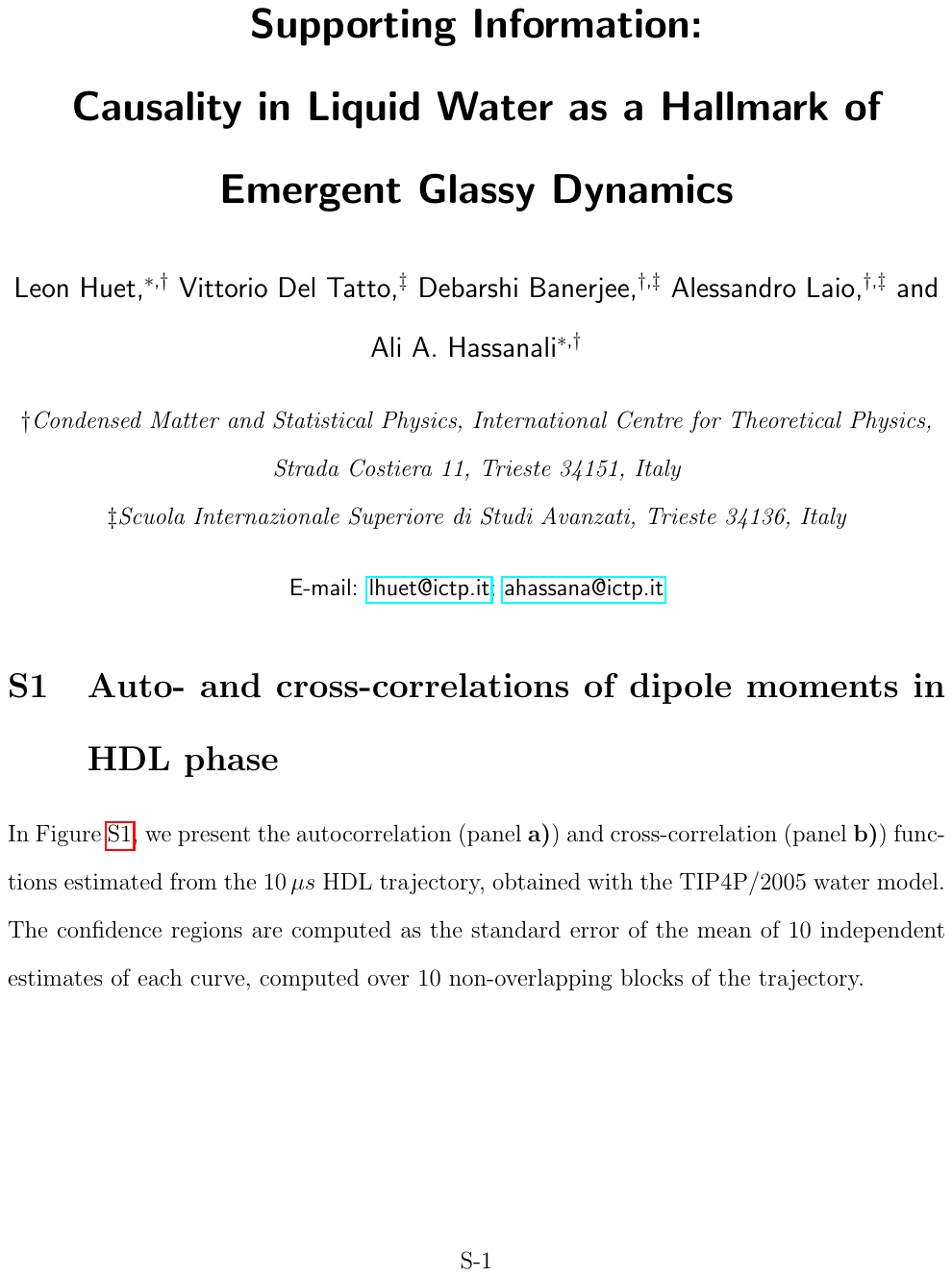}

\end{document}